\documentstyle[12pt]{article} \tolerance=10000
\begin{document}
\vskip .7cm
\begin{center}
{\bf { New topological field theories in two dimensions}}

\vskip 2cm

{\bf R. P. Malik}
\footnote{ E-mail: malik@boson.bose.res.in  }\\
{\it S. N. Bose National Centre for Basic Sciences,} \\
{\it Block-JD, Sector-III, Salt Lake, Calcutta-700 098, India} \\

\vskip 2.5cm

\end{center}
{\bf Abstract.}               
It is shown that two$(1 + 1)$-dimensional (2D) free Abelian- and 
self-interacting non-Abelian gauge theories (without any interaction with 
matter fields) belong to a new class of topological field theories. These new
theories capture together some of the key features of Witten- and Schwarz type 
of topological field theories because they are endowed with symmetries that are
reminiscent of the Schwarz type theories but their Lagrangian density has the
appearance of the  Witten type theories.  The topological invariants for these 
theories are computed on a 2D compact manifold and their recursion relations 
are obtained. These new theories are shown to provide a class of tractable 
field theoretical models for the Hodge theory in two dimensions of flat 
(Minkowski) spacetime where there are no propagating degrees of freedom 
associated with the 2D gauge boson. 

\baselineskip=16pt

\vskip 1cm

\newpage

\noindent
{\bf 1 Introduction}\\

\noindent
The history of modern developments in theoretical high energy physics
is rich with many cardinal examples which have provided a meeting-ground for
theoretical physicists as well as mathematicians.
One such example is the subject of topological field theories (TFTs)
which has encompassed in its folds such diverse areas of theoretical physics 
and mathematics as Chern-Simon theories, string theories and
matrix models, two-dimensional
topological gravity, Morse theory, Donaldson- and Jones polynomials, etc. 
(see, e.g., Refs. [1--12] and references therein). Broadly speaking, there
are two types of TFTs. Witten type TFTs [2,3] are the ones where the quantum
action (or the Lagrangian density (${\cal L}_{W}$) itself) can be written
as the Becchi-Rouet-Stora-Tyutin (BRST) (anti)commutator, i.e.,
$$
\begin{array}{lcl}
{\cal L}_{W} = \{ Q_{B}^{(w)}, V(\Phi, g)\}
\end{array}\eqno(1.1)
$$
where $Q_{B}^{(w)}$ is the conserved 
($\dot Q_{B}^{(w)} = 0$) and nilpotent ($ (Q_{B}^{(w)})^2 = 0$)
BRST charge (in general, metric independent) and $ V (\Phi, g)$ is a local
expression for the field operators as a function of generic field $\Phi$
and the metric $g$ of spacetime manifold on which the theory
is defined. Here the BRST charge is constructed by combining a topological 
shift symmetry with some kind of local gauge symmetries. In contrast,
for the Schwarz type of TFTs [1], the classical action is metric independent
and the sum of gauge-fixing and Faddeev-Popov ghost terms are BRST
(anti)commutator. In the language of Lagrangian density 
($ {\cal L}_{S}$), the Schwarz type of TFTs bear an outlook as 
$$
\begin{array}{lcl}
{\cal L}_{S} = {\cal L}_{C} + \{ Q_{B}^{(s)}, S(\Phi, g)\}
\end{array}\eqno(1.2)
$$
where ${\cal L}_{C}$ is the classical metric-independent Lagrangian density
that cannot be expressed as the BRST (anti)commutator  and $S(\Phi, g)$
is a local field operator that contains all the metric dependence of the 
theory. For the Schwarz type TFTs, the BRST charge $Q_{B}^{(s)}$ generates
only some local gauge symmetries (and there is no presence of a subtle local
topological shift symmetry, which is a characteristic feature of Witten
type TFTs). It is now obvious that, for both types of TFTs, the symmetric 
energy momentum tensor is a BRST (anti)commutator. This entails the
Hamiltonian density of the theory to be a BRST (anti)commutator. As a
consequence, there are no energy excitations in the physical sector of the
theory as all the physical states are BRST invariant (i.e. 
$Q_{B}^{(w, s)} |phys> = 0$). This also ensures that all the correlation
functions of the observables (i.e. BRST invariant operators) for the theory 
are independent of the choice of the metric on the spacetime manifolds.

For the case of a manifold with a trivial flat metric, the TFTs are 
those theories where
there are no propagating (dynamical) degrees of freedom
associated with the fields (see, e.g., Ref. [12]). In the present
paper, we shall concentrate on the free Abelian- and self-interacting
non-Abelian gauge theories in two-dimensions of spacetime (endowed with a flat
Minkowski metric) and show that there are no propagating degrees of freedom 
associated with the gauge bosons of these theories because of the presence of
two nilpotent charges which are required to have consistency with the Hodge
decomposition theorem (HDT). The symmetries, corresponding to these charges, 
gauge out
the dynamical degrees of freedom of the gauge bosons and the theory becomes
topological in nature. In fact, these nilpotent charges will be 
shown to be analogous to the exterior derivative $d$ ($d^2 = 0$) and
co-exterior derivative $\delta$ ($\delta = \pm * d *, \delta^2 = 0$) of
differential geometry which are required in the definition of the HDT
which states that, on a compact manifold, any arbitrary  $n$-form 
$f_{n} (n = 0, 1, 2, ...)$ can be written as the unique sum of a harmonic
form $h_{n}$ ($ \Delta h_{n} = 0, d h_{n} = 0, \delta h_{n} = 0$), an
exact form $(d e_{n-1})$ and a co-exact form $(\delta  c_{n+1})$ as [13-17]
$$
\begin{array}{lcl}
f_{n} = h_{n} + d e_{n-1} + \delta c_{n+1}
\end{array}\eqno(1.3)
$$
where $\Delta = (d + \delta)^2 = d \delta + \delta d$ is the Laplacian
operator. The set of operators ($d , \delta, \Delta$) is called the de Rham
cohomology operators of differential geometry as they define the cohomological
properties of a given differential form on a compact manifold.

It has been a long-standing problem to express the de Rham cohomology operators
in the language of some {\it local} symmetry properties of a given Lagrangian 
density. Normally, $d (d^2 = 0)$ operator is identified with the local
BRST charge $Q_{B} (Q_{B}^2 = 0)$ which generates a local, continuous, 
covariant and nilpotent symmetry transformation for a BRST invariant Lagrangian
density corresponding to a given gauge theory. Some very interesting and
enlightening attempts [18-21] have been made to express 
$\delta$ and $\Delta$ for the interacting
(non)Abelian gauge theories in arbitrary spacetime dimension but the symmetry
transformations turn out to be nonlocal and noncovariant. In the covariant
formulation, nilpotency is achieved only for some
specific values of the parameters
of the theory [22]. Recently, however, it has been shown [23-25] that 2D free
Abelian- and self-interacting non-Abelian gauge theories provide a couple of
field theoretical models for the Hodge theory where all the de Rham cohomology
operators $(d, \delta, \Delta)$ correspond to local and conserved charges
which generate local, continuous, covariant and nilpotent (for $d$ and $\delta$)
symmetries for the BRST invariant Lagrangian density of these theories.
In fact, the BRST symmetry (analogue of $d$) corresponds to a transformation
in which the kinetic energy terms of these theories remain invariant. On the
other hand, co-BRST symmetry (analogue of $\delta$) is found to be a symmetry
transformation under which the gauge-fixing terms
\footnote{ The one-form $A = A_{\mu} dx^\mu$ defines the vector
potential for the Abelian gauge theory. The zero-form (gauge-fixing) 
$\delta A = (\partial \cdot A)$ and the curvature two-form 
(field strength tensor) $F^{(A)} = d A$ are 
`Hodge dual' to each-other in any arbitrary dimension of spacetime. Here 
$\delta = \pm * d *$ is the co-exterior derivative w.r.t. $d$. The same
is not true ($ F^{(N)} \neq d A$) for the non-Abelian gauge theory
(see, e.g., Ref. [14]).}  
remain invariant. The anticommutator of these transformations (analogue of 
$\Delta$) leaves the Faddeev-Popov ghost terms invariant.  In these attempts, 
the topological features of these theories have been very briefly mentioned. The 
central aim of our present paper is to apply the insights gained in our
earlier studies [23-25] to furnish an elaborate proof of the topological nature
of these theories as they capture in their realm some very interesting 
{\it new} features. For instance, the form of the Lagrangian
density for these theories turns out to be like Witten type TFT but the
underlying symmetries are found to be of Schwarz type. Furthermore,         
there are four sets of topological invariants for these theories. These 
are computed on a 2D compact manifold and energy-momentum tensor is shown 
to be the sum of a BRST- and co-BRST  anticommutator. By exploiting the HDT,
it is demonstrated that there are no energy excitations in the physical sector.

The outline of our present paper is as follows.

In section 2, we set up the
notations and recapitulate the essentials of our earlier work [23,24]
so that the paper can be self-contained. Here we show that the
2D free Abelian gauge theory is a topological field theory by exploiting
the basic ingredients of BRST cohomology and HDT.
We demonstrate further that this free theory is also a perfect
example of a Hodge theory where, not only the cohomological operators
($d, \delta, \Delta$) are expressed in terms of generators for some
{\it local} symmetries, but even the Hodge duality $(*)$ operation is 
shown to correspond to the existence of
a couple of discrete symmetries in the theory. With
respect to four conserved and nilpotent charges of the theory, we
derive four sets of topological invariants which are shown to be
inter-related by the Hodge duality $(*)$ operation and the presence of a 
discrete symmetry for the ghost action. In fact, the requirement of a specific
relationship between the set of topological invariants w.r.t. (anti)BRST-
and (anti)co-BRST charges,
singles out  one of the two discrete symmetries of the Lagrangian density
which are the analogue of Hodge $(*)$ operation.

Section 3 is devoted to the discussion of a self-interacting 2D non-Abelian
gauge theory (without any interaction with matter fields). 
We derive all the four sets of topological invariants
on a 2D compact manifold (w.r.t. all the conserved and nilpotent charges 
in the theory) and obtain their recursion relations. In analogy
with the Abelian gauge theory, we derive a discrete symmetry as
an analogue of the Hodge $(*)$ operation for the non-Abelian gauge theory by
requiring a certain specific transformation property 
for the topological invariants
of the theory. This discrete symmetry reduces to its Abelian counterpart
(as Hodge $(*)$ operation) in the limit when the coupling constant $g$ of 
non-Abelian gauge theory goes to zero ($ g \rightarrow 0$).

Finally, in section 4, we make some concluding remarks and point out
some future directions that can be pursued for further extension of our
work. \\

\noindent
{\bf 2 Abelian gauge theory}\\

\noindent
Let us begin with a two$(1 + 1)$-dimensional
\footnote{ We adopt the notations in which the flat 2D Minkowski metric
$\eta_{\mu\nu} = $ diag$\; (+1, -1)$  and anti-symmetric Levi-Civita tensor
$\varepsilon_{\mu\nu} \varepsilon^{\nu \lambda}
= \delta^\lambda_{\mu}, \varepsilon^{\mu\nu} \varepsilon_{\mu\nu} = - 2!,  
\varepsilon_{01} = \varepsilon^{10}
= + 1, \;F_{01} = \partial_{0} A_{1} - \partial_{1} A_{0} 
= E = - \varepsilon^{\mu\nu} \partial_{\mu} A_{\nu} = F^{10},
\Box = \eta^{\mu\nu} \partial_{\mu} \partial_{\nu} = (\partial_{0})^2
- (\partial_{1})^2, \;\dot f = \partial_{0} f $. 
Note that there is no magnetic component in the 2D field strength
tensor $F_{\mu\nu}$. Here Greek indices: 
$\mu, \nu, \lambda.. = 0, 1$ stand for the 
Minkowski flat spacetime directions.}
BRST invariant Lagrangian density (${\cal L}_{b}$) for the free Abelian 
gauge theory in the Feynman gauge [26-28]
$$
\begin{array}{lcl}
{\cal L}_{b} = - \frac{1}{4} F^{\mu\nu} F_{\mu\nu}
- \frac{1}{2} (\partial \cdot A)^2  - i \partial_{\mu} \bar C
\partial^{\mu} C \end{array}\eqno (2.1a)
$$
$$
\begin{array}{lcl}
{\cal L}_{b} =  \frac{1}{2} E^2
- \frac{1}{2} (\partial \cdot A)^2  - i \partial_{\mu} \bar C
\partial^{\mu} C \end{array}\eqno (2.1b)
$$
where $F_{\mu\nu} = \partial_{\mu} A_{\nu} - \partial_{\nu} A_{\mu} $
is the curvature tensor derived from the two-form $F = d A$, 
$(\partial \cdot A)$ is the gauge-fixing term derived from the zero-form
$(\partial \cdot A) = \delta A$,
$(\bar C)C$ are the Faddeev-Popov (anti)ghost fields ($ \bar C^2 
= C^2 = 0$) and indices $\mu,\nu = 0, 1$ represent the flat Minkowski time
and space directions. It has been shown [23,24] that the above 
Lagrangian density
remains quasi-invariant (i.e. $\delta_{b} {\cal L}_{b} = - \eta \partial_{\mu}
[(\partial \cdot A) \partial^{\mu} C],
\; \delta_{d} {\cal L}_{b} = \eta \partial_{\mu}
[ E \partial^\mu \bar C]$) under the following on-shell 
$(\Box C = 0, \Box \bar C = 0) $ nilpotent
($ \delta_{b}^2 = 0, \delta_{d}^2 = 0$) BRST
($\delta_{b}$)- and dual BRST ($\delta_{d}$) transformations:    
$$
\begin{array}{lcl}
\delta_{b} A_{\mu} &=& \eta\; \partial_{\mu}\; C \;\;\;\;\;\qquad\;\;\;\;\;
\delta_{d} A_{\mu} = - \eta \varepsilon_{\mu\nu} \partial^{\nu} \bar C
\nonumber\\
\delta_{b} C &=& 0 \;\;\;\;\;\;\;\;\;\qquad \;\;\;\; \;\;\;\;\;
\delta_{d} \bar C  = 0 \nonumber\\
\delta_{b} \bar C &=& - i \;\eta\; (\partial \cdot A) \; \qquad \;
\delta_{d} C = - i \;\eta \;E \nonumber\\
\delta_{b} E &=& 0 \;\;\;\;\; \;\;\;\;\qquad \;\;\;\;\;\;\;\;\;
\delta_{d} (\partial \cdot A) = 0
\end{array}\eqno (2.2)
$$
where $\eta$ is an anticommuting $(\eta C = - C \eta, \eta \bar C = - \bar C
\eta)$ spacetime independent transformation parameter. It will be noticed that,
under $\delta_{b}$, it is the electric field $E$ (derived by the application
of $d$ on one-form $A = A_{\mu} dx^\mu$) that remains invariant and the
gauge-fixing term transforms (to compensate for the term coming from the 
variation of ghost
term). In contrast, under $\delta_{d}$, it is the gauge-fixing term 
$(\partial\cdot A)$ (derived from one-form $A = A_{\mu} dx^\mu$ by 
the application of $\delta$) that remains invariant and 
the electric field transforms. The conserved, {\it local} and nilpotent
($ Q_{b}^2 = 0, Q_{d}^2 = 0$) generators for the above transformations are
$$
\begin{array}{lcl}
Q_{b} = {\displaystyle \int} d x\; \bigl [\; \partial_{0} (\partial \cdot A) C 
- (\partial \cdot A) \dot C \;\bigr ] \qquad Q_{d} = {\displaystyle \int} d x 
\;\bigl [ \;E \dot {\bar C} - \dot E \bar C \;\bigr ].
\end{array} \eqno(2.3)
$$
It is very natural to expect that the anticommutator of these two
transformations ($\{ \delta_{b}, \delta_{d} \} = \delta_{w}$) would
also be the symmetry transformation ($\delta_{w}$) for the Lagrangian
density. This is indeed the case as can be seen
that under the following bosonic ($ \kappa = - i \eta \eta^{\prime}$)
transformations
$$
\begin{array}{lcl}
\delta_{w} A_{\mu} &=& \kappa (\partial_{\mu} E
- \varepsilon_{\mu\nu} \partial^{\nu} (\partial \cdot A)) \qquad\;
\delta_{w} (\partial \cdot A) = \kappa \Box E 
\nonumber\\
\delta_{w} E &=&  \kappa \Box (\partial \cdot A) \;\qquad \;
\delta_{w} C = 0 \;\qquad\; \delta_{w} \bar C = 0
\end{array}\eqno(2.4)
$$
the Lagrangian density (2.1) transforms as: $ \delta_{w} {\cal L}_{b}
= \kappa \partial\; [\; E \partial^\mu (\partial \cdot A) - (\partial \cdot A)
\partial^\mu E \;]$. Here $\eta$ and $\eta^\prime$ are the transformation
parameters corresponding to $\delta_{b}$ and $\delta_{d}$ respectively. The
generator for the above transformation is
$$
\begin{array}{lcl}
W = {\displaystyle \int} dx\; \bigl [\; \partial_{0} (\partial \cdot A) E
- (\partial_{0} E) (\partial \cdot A)\;\bigr ].
\end{array}\eqno(2.5)
$$
The global scale invariance of the Lagrangian density (2.1)
under $ C \rightarrow e^{- \Sigma} C, \bar C \rightarrow e^{\Sigma}
\bar C, A_{\mu} \rightarrow A_{\mu}$, (where $\Sigma$
is a global parameter), leads to the derivation of a conserved ghost charge
($Q_{g}$)
$$
\begin{array}{lcl}
Q_{g} = - i {\displaystyle \int} d x \;\bigl [\; C \dot {\bar C} 
+ \bar C \dot  C \; \bigr ].
\end{array} \eqno(2.6)
$$
Together, these conserved charges satisfy the following algebra [23,24]
$$
\begin{array}{lcl}
&& [ W, Q_{k} ] = 0 \;\;\;\qquad\;\;\; k = g, b, d, ab, ad \nonumber\\
&& Q_{b}^2 = Q_{d}^2 = Q_{ab}^2 = Q_{ad}^2 = 0 \qquad
\{ Q_{d}, Q_{ad} \} = 0 \nonumber\\
&& \{ Q_{b}, Q_{d} \} = \{Q_{ab}, Q_{ad} \} = W \qquad
\{ Q_{b}, Q_{ab} \} = 0 \nonumber\\
&& i [ Q_{g}, Q_{b} ] = Q_{b} \qquad i [ Q_{g}, Q_{ab} ] = - Q_{ab}
\nonumber\\
&& i [ Q_{g}, Q_{d} ] = - Q_{d} \qquad i [ Q_{g}, Q_{ad} ] = Q_{ad}
\end{array}\eqno(2.7)
$$
where $Q_{ab}$ and $Q_{ad}$ are the anti-BRST- and anti-dual BRST 
charges which can be readily obtained from (2.3)
by the replacement :$ C \rightarrow \pm i \bar C, \bar C \rightarrow \pm i C$ 
\footnote {Note that the discrete transformations:
$ C \rightarrow \pm i \bar C, \bar C \rightarrow \pm i C$ 
are the symmetry transformations for the ghost action ($ I_{F.P.} 
= - i \int d^{D} x\; \partial_{\mu} \bar C \partial^{\mu} C$)
in any arbitrary dimension of spacetime.}. It can be seen that the ghost
number for $Q_{b}$ and $Q_{ad}$ is $+1$ and that of $Q_{d}$ and $Q_{ab}$
is $-1$. Now, given a state $|\phi>$ (with ghost number $n$) in the quantum
Hilbert space (i.e. $i Q_{g} |\phi> = n |\phi>$), it can be readily seen, using
the above algebra (2.7), that
$$
\begin{array}{lcl}
i Q_{g} Q_{b} | \phi > &=& (n + 1) Q_{b} |\phi> 
\qquad i Q_{g} Q_{ad} | \phi > = (n + 1) Q_{ad} | \phi > \nonumber\\
i Q_{g} Q_{d} | \phi > &=& (n - 1) Q_{d} |\phi>
\qquad i Q_{g} Q_{ab} | \phi > = (n - 1) Q_{ab} | \phi > \nonumber\\
i Q_{g} W   \;| \phi > &=& n \;    W \;  |\phi>.
\end{array}\eqno(2.8)
$$   
This shows that the ghost numbers for the 
states $Q_{b} |\phi>$ (or $ Q_{ad} | \phi >$), $Q_{d} |\phi>$ (or
$ Q_{ab} | \phi > $) and $W |\phi>$ in the quantum Hilbert space 
are $ (n + 1), (n - 1)$ and $ n $ respectively. As far as underlying algebra 
is concerned, the above symmetry generators $Q_{b}, Q_{d}$ and $W$ obey
the same kind of algebra as their counterparts (de Rham 
cohomology operators $ d, \delta$ and $ \Delta$) in differential geometry.
The latter algebra can be succinctly expressed as 
$$
\begin{array}{lcl}
&& d^2 = 0 \qquad  \delta^2 = 0 \qquad \Delta =
(d + \delta)^ 2 = d \delta + \delta d \nonumber\\
&& [ \Delta , d ] = 0 \qquad  [\Delta , \delta ] = 0 \qquad
\Delta = \{ d, \delta \} \neq 0.
\end{array}\eqno(2.9a)
$$
It is a peculiarity
of the BRST formalism that the above cohomological operators can be also 
identified with the generators $ Q_{ad}, Q_{ab}$ and $W 
= \{Q_{ab}, Q_{ad} \}$ respectively. Thus, the mapping is: $ (Q_{b}, Q_{ad})
\Leftrightarrow d, \; (Q_{d}, Q_{ab}) \Leftrightarrow \delta, \; W 
= \{ Q_{b}, Q_{d} \} = \{ Q_{ad}, Q_{ab} \} \Leftrightarrow \Delta$. This
analogy enables us to express the Hodge decomposition theorem in the quantum
Hilbert space of states where any arbitrary state $|\phi>_{n}$ (with ghost 
number $n$) can be written as the sum of a harmonic state $|\omega>_{n}$ ($
W |\omega>_{n} = 0, Q_{b} |\omega>_{n} = 0, Q_{d} |\omega>_{n} = 0$), a BRST
exact state ($Q_{b} |\theta>_{n-1}$) and a co-BRST exact state ($Q_{d} 
|\chi>_{n+1}$). Mathematically, this statement (which is the analogue of
eqn. (1.3)) can be expressed, in two equivalent ways, as
$$
\begin{array}{lcl}
|\phi>_{n} = |\omega>_{n} + Q_{b} |\theta>_{n-1} + Q_{d} |\chi>_{n+1}
\equiv  |\omega>_{n} + Q_{ad} |\theta>_{n-1} + Q_{ab} |\chi>_{n+1}.
\end{array}\eqno(2.9b)
$$

It is worth pointing out that the sets of charges $(Q_{b}, Q_{d})$ and
$(Q_{ab}, Q_{ad})$) are the dual sets, because $Q_{b}$ and $Q_{d}$ are dual to 
each-other as are $Q_{ab}$ and $Q_{ad}$. To elaborate this claim, it can
be seen that, under the following separate and independent transformations 
$$
\begin{array}{lcl}
C \rightarrow \pm i \bar C \quad \bar C \rightarrow \pm i C \quad
A_{\mu} \rightarrow A_{\mu} \quad \partial_{\mu} \rightarrow
\pm i \varepsilon_{\mu\nu} \partial^\nu
\end{array}\eqno(2.10)
$$
$$
\begin{array}{lcl}
C \rightarrow \pm i \bar C \quad \bar C \rightarrow \pm i C \quad
A_{\mu} \rightarrow \mp i \varepsilon_{\mu\nu} A^{\nu} 
\end{array}\eqno(2.11)
$$
the two-form (electric) field $E$ and the zero-form (gauge-fixing) 
field $(\partial \cdot A)$ are
related with each-other as: $ E \rightarrow \pm i (\partial \cdot A),\;
(\partial \cdot A) \rightarrow \pm i E$. Thus, we see that under 
the above transformations: (i) the Lagrangian density (2.1) remains invariant.
(ii) The dual BRST symmetry transformations $\delta_{d}$ can be obtained from
the BRST transformations $\delta_{b}$ in (2.2). (iii) The symmetry generators
$Q_{k}, (k = b, ab, d, ad, g)$ and $W$ transform as: 
$$
\begin{array}{lcl}
&& Q_{b} \rightarrow Q_{d} \; \quad
Q_{d} \rightarrow Q_{b} \quad Q_{ab} \rightarrow Q_{ad} 
\nonumber\\
&& Q_{ad} \rightarrow Q_{ab} \quad  Q_{g} \rightarrow - Q_{g}
\quad W \rightarrow W.
\end{array}\eqno(2.12)
$$
(iv) The algebraic structure of (2.7) remains form-invariant under (2.12).

The transformations (2.10) and (2.11) are the
analogue of the Hodge $(*)$ operation of differential geometry. To clarify 
this assertion, first, we note the  consequences of two successive operations 
of $(*)$ on the generic field $\Phi$ of the theory, namely;
$$
\begin{array}{lcl}
*\; \bigl (\; *\; \Phi \bigr ) = \pm \; \Phi.
\end{array}\eqno(2.13)
$$
Here $*$ operation corresponds to 
transformations (2.10) and $(+)$ sign stands for
$\Phi = A_{\mu}$ and  $(-)$ sign for $ \Phi = C, \bar C, E, (\partial \cdot A)$.
Under transformations (2.11), the analogue of (2.13) is 
$ *\; (*\; \Phi) = - \Phi$
for all the fields of the theory (i.e. $ \Phi = A_{\mu}, C, \bar C, E,
(\partial \cdot A)$. Now, it is straightforward to check that $\delta_{d}$
and $\delta_{b}$ are related to each other as
$$
\begin{array}{lcl}
\delta_{d} \; \Phi = \pm \;*\; \delta_{b} \; *\; \Phi
\end{array}\eqno(2.14)
$$
where $\delta_{d}$ and $\delta_{b}$ are the nilpotent transformations in (2.2)
and the ($*$) operation
 corresponds to transformations in (2.10). The signs in (2.14) are
governed by the corresponding signatures in (2.13). For the ($*$) operation 
corresponding to (2.11), the analogue of (2.14), is
$$
\begin{array}{lcl}
\delta_{d} \; \Phi = - \;*\; \delta_{b} \; *\; \Phi
\end{array}\eqno(2.15)
$$
for the generic field $\Phi = A_{\mu}, C, \bar C, E, (\partial \cdot A)$. 
It is obvious that the relation between nilpotent transformations $\delta_{d}$
and $\delta_{b}$, acting on a generic field $\Phi$, is same as the relation
between dual exterior derivative $\delta (= \pm * d *)$ and exterior derivative
$d$ acting on a differential form defined on a compact manifold. It
will be noticed that duality transformations in 2D and 4D are different [29-31].
This is the reason that, for the 4D $(3 + 1)$ theories, it has been shown 
[32] that under $*$ operation: $ Q_{b} \rightarrow Q_{d}, Q_{d} \rightarrow 
- Q_{b}$ which is like the electromagnetic duality transformations for the 
Maxwell equations where: $ {\bf E} \rightarrow {\bf B}, {\bf B} 
\rightarrow - {\bf E}$ . In fact, it is due to 
the peculiarity of duality transformations in 2D that a
reverse relation also exist which allows one to express 
$\delta_{b}$ in terms of $\delta_{d}$ as
$$
\begin{array}{lcl}
\delta_{b} \; \Phi = \pm \;*\; \delta_{d} \; *\; \Phi 
\qquad \mbox {and} \qquad
\delta_{b} \; \Phi = - \;*\; \delta_{d} \; *\; \Phi
\end{array}\eqno(2.16)
$$
corresponding to transformations (2.10) and (2.11), respectively.

Exploiting the fact that conserved charges $Q_{r} (r = b, d, ab, ad)$ are
the generators for the transformations $\delta_{r}\; \Phi = - i \;
\eta [\;\Phi, Q_{r}\;]_{\pm}$ where $(+)-$ stand for the (anti)commutator
corresponding to $\Phi$ being (fermionic)bosonic in nature, it can be readily
seen that the Lagrangian density in (2.1) can be written, modulo some total
derivatives, as
$$
\begin{array}{lcl}
{\cal L}_{b} = \{ Q_{d}, T_{1}\} 
+ \{ Q_{b}, T_{2}\} \equiv
\{ Q_{ad}, P_{1}\} + \{ Q_{ab}, P_{2}\}
\end{array}\eqno(2.17)
$$ 
where $T_{1} = \frac{1}{2}\; E C, \; T_{2} = - \frac{1}{2} (\partial \cdot A)
\bar C, \;P_{1} = \frac{i}{2}\; E \bar C,\;  P_{2} = - \frac{i}{2}
(\partial \cdot A) C$. Furthermore, using the on-shell nilpotent symmetries
of (2.2), it can be checked that the above Lagrangian density can be 
re-expressed as a sum of the (anti)BRST- and  (anti)-dual BRST invariant parts
and a total derivative, as
$$
\begin{array}{lcl}
\eta {\cal L}_{b} =  \delta_{d} \;(iT_{1}) + \delta_{b}\; (i T_{2}) 
+ \eta \;\partial_{\mu} Y^\mu \equiv
\delta _{ad}\; (i P_{1}) + \delta_{ab}\; (i P_{2}) + \eta \;
\partial_{\mu} Y^\mu
\end{array}\eqno(2.18)
$$ 
where the nilpotent transformations $\delta_{ab}$ and $\delta_{ad}$ can
be readily derived from (2.2) by exploiting the substitution: $ C \rightarrow 
i \bar C, \;\bar C \rightarrow  i C$ and $Y^\mu = \frac{i}{2}
(\bar C \partial^\mu C + \partial^\mu \bar C  C)$. The appearance of the
Lagrangian  density (2.1) (in the form (2.17)) is reminiscent of the Witten-type
topological field theories (1.1) where it is possible to express the Lagrangian
density of a TFT as a BRST (anti)commutator. Even though in our case, we have
two sets of nilpotent charges $(Q_{b}, Q_{d})$ as well as $(Q_{ab},Q_{ad})$, 
the outlook of the Lagrangian
density (2.17) is same as the Witten type TFTs because the physical states are
the harmonic states (of the Hodge decomposition theorem) which satisfy
$Q_{(b, ab)} |phys> = 0, Q_{(d, ad)} |phys> = 0$. It should be noted that 
the appearance             
in (2.17), is completely different from the Schwarz type of theories where
the Lagrangian density (1.2) is a sum of a BRST (anti)commutator and a piece 
that can never be expressed as a BRST (anti)commutator [1,12]. At this stage,
however, we note that we have only local gauge
type symmetries and there is no trace of any topological shift 
symmetries. Hence, it is clear that, from the symmetry point of view, the
free 2D $U(1)$ gauge theory is like Schwarz type topological theories.

One of the key properties of TFTs is the absence of any energy excitations
in the theory. This aspect is governed by the expression for the symmetric 
energy-momentum tensor $(T_{\alpha \beta}^{(s)})$. It is 
interesting to check that
the expression for this symmetric tensor  for the generic field
$\Phi = A_{\mu}, C, \bar C$, present in the Lagrangian density (2.1), is
$$
\begin{array}{lcl}
T^{(s)}_{\alpha\beta} &=& \frac{1}{2} \;\partial_{\alpha} \Phi\;
{\displaystyle \frac {\partial {\cal L}_{b}} {\partial_{\beta} \Phi}}
+ \frac{1}{2} \;\partial_{\beta} \Phi\;
{\displaystyle \frac {\partial {\cal L}_{b}} {\partial_{\alpha} \Phi}}
- \eta_{\alpha\beta} \;{\cal L}_{b} \nonumber\\
&\equiv&  
- \frac{1}{2}\; [\; \varepsilon_{\alpha \rho} E + \eta_{\alpha \rho}
(\partial \cdot A)\; ]\; \partial_{\beta} A^\rho
- \frac{1}{2} \;[\; \varepsilon_{\beta \rho} E + \eta_{\beta \rho}
(\partial \cdot A)\; ]\; \partial_{\alpha} A^\rho \nonumber\\
&-& i \partial_{\alpha} \bar C \partial_{\beta} C 
- i \partial_{\beta} \bar C \partial_{\alpha} C - \eta_{\alpha \beta}
\;{\cal L}_{b}.
\end{array}\eqno(2.19a)
$$
This equation, with the use of (2.17), can be explicitly expressed as
\footnote{Here, and in what follows, we shall be exploiting only nilpotent
charges $Q_{b}$ and $Q_{d}$ for our purposes. However, $Q_{ab}$ and $Q_{ad}$
could be used equally well for the same objectives. All one has to do is to 
exploit the substitution: $C \rightarrow \pm i \bar C, \bar C \rightarrow 
\pm i C$ judiciously.}  
$$
\begin{array}{lcl}
T^{(s)}_{\alpha\beta} 
&=& \{ Q_{b}, V^{(1)}_{\alpha\beta} \} + \{ Q_{d}, V^{(2)}_{\alpha\beta} \}
\nonumber\\
V^{(1)}_{\alpha\beta} &=& \frac{1}{2}\; \bigl [\;
(\partial_{\alpha} \bar C) A_{\beta} +
(\partial_{\beta} \bar C) A_{\alpha} + \eta_{\alpha\beta} (\partial \cdot A)
\bar C \;\bigr ] \nonumber\\
V^{(2)}_{\alpha\beta} &=& \frac{1}{2}\; \bigl [\;
(\partial_{\alpha} C) \varepsilon_{\beta\rho} A^{\rho } +
(\partial_{\beta} C) \varepsilon_{\alpha\rho} A^{\rho } 
- \eta_{\alpha\beta} E C \;\bigr ]. 
\end{array}\eqno(2.19b)
$$
This shows that, when the Hamiltonian density $\hat T^{(s)}_{00}$ is 
sandwiched between two physical states (i.e. $< phys | \hat T_{00}^{(s)} 
| phys^{\prime} > = 0$) it turns out to be zero because hermitian operators 
$Q_{b}, Q_{d}$  annihilate the harmonic states (which are the BRST- 
and co-BRST invariant {\it physical states} 
($ Q_{b} |phys> = 0, Q_{d} |phys> = 0$) in the theory). In fact, conditions
$Q_{b} |phys> = 0, Q_{d} |phys> = 0$ imply that $(\partial \cdot A) |phys>
= 0, \varepsilon^{\mu\nu} \partial_{\mu} A_{\nu} |phys> = 0$, respectively
[23,24].
This ensures that there are no propagating degrees of freedom in the theory
as both the components $A_{0}$ and $A_{1}$ of a 2D photon are conserved
quantities (w.r.t. time). In other words, there is no evolution in the
system w.r.t. the evolution (time) parameter of the theory. This condition 
confirms the topological nature of the free
2D $U(1)$ gauge theory in the flat Minkowski spacetime.

The topological nature is further confirmed 
by the existence of two sets of topological invariants w.r.t. 
conserved and on-shell ($ \Box C = \Box \bar C = 0$) nilpotent 
($ Q_{b}^2 = 0, \; Q_{d}^2 = 0$) BRST- and co-BRST charges. For the 2D
compact manifold, these are
$$
\begin{array}{lcl}
I_{k} = {\displaystyle \oint}_{C_{k}}\; V_{k} \qquad
J_{k} = {\displaystyle \oint}_{C_{k}}\; W_{k} \qquad (k = 0, 1, 2)
\end{array}\eqno(2.20)
$$
where $C_{k}$ are the $k$-dimensional homology cycles in the 2D manifold
and $V_{k}$ and $W_{k}$ are the $k$-forms. These forms, w.r.t. the BRST
charge $Q_{b}$, are
$$
\begin{array}{lcl}
V_{0} &=& - (\partial \cdot A) C \qquad 
V_{1} = \bigl [\; - (\partial \cdot A) A_{\mu} 
+ i C \partial_{\mu} \bar C \;\bigr ]\; dx^\mu \nonumber\\
V_{2} &=& i \;\bigl [\;A_\mu \partial_\nu \bar C - \frac{\bar C}{2}
F_{\mu\nu} \; \bigr ]\; dx^\mu \wedge dx^\nu
\end{array}\eqno(2.21)
$$
and the same, w.r.t. the dual BRST charge $Q_{d}$, are
$$
\begin{array}{lcl}
W_{0} &=& E \bar C \qquad 
W_{1} = \bigl [\; \bar C \varepsilon_{\mu\rho} \partial^\rho C 
- i E A_{\mu} \; \bigr ] \;dx^\mu \nonumber\\
W_{2} &=& i \bigl [\;\varepsilon_{\mu\rho} \partial^\rho C A_\nu
+ \frac{C}{2}  \varepsilon_{\mu\nu} (\partial \cdot A) \;\bigr ]\;
dx^\mu \wedge dx^\nu. 
\end{array}\eqno(2.22)
$$
It will be noticed here that there are two more sets of topological invariants
$(\tilde V_{k}, \tilde W_{k}$)
w.r.t. the conserved and on-shell ($\Box C = \Box \bar C = 0$) nilpotent 
($Q_{ab}^2 = Q_{ad}^2 = 0$) anti-BRST- and anti-dual BRST charges.  These can
be derived from (2.21) and (2.22) by the substitution: $C \rightarrow i \bar C,
\; \bar C \rightarrow i C$. For $k  = 1, 2$, all these four invariants obey a 
specific recursion relation 
$$
\begin{array}{lcl}
\delta_{b} \;V_{k} &=& \eta \;d \; V_{k - 1} \qquad
\delta_{ab}\; \tilde V_{k} = \eta \;d \; \tilde V_{k - 1} \qquad
d = dx^\mu \; \partial_{\mu} \nonumber\\
\delta_{d} \; W_{k} &=& \eta \;\delta\;  W_{k-1} 
\qquad \delta_{ad} \;\tilde W_{k} = \eta \;\delta \;\tilde W_{k-1}
\qquad \delta = i \;dx^\mu \; \varepsilon_{\mu\nu}\; \partial^{\nu} 
\end{array}\eqno(2.23)
$$
which is a typical feature for the existence of any TFTs. It is very 
interesting to note that,
under $(*)$ operation corresponding to transformations in (2.10), the above
topological invariants, for $k = 0, 1, 2$, transform as
$$
\begin{array}{lcl}
 V_{k} \rightarrow W_{k}\; \quad
\tilde V_{k} \rightarrow \tilde W_{k} \quad  
 W_{k} \rightarrow (-1)^k \; V_{k} 
\quad  \tilde W_{k} \rightarrow (-1)^k\; \tilde V_{k}.
\end{array}\eqno(2.24)
$$
Mathematically, this statement can be succinctly expressed as
$$
\begin{array}{lcl}
*\; V_{k} = W_{k} \quad *\; \tilde V_{k} = \tilde W_{k}  \quad 
*\; \bigl ( *\; V_{k} \bigr ) = (-1)^k\; V_{k} \quad
*\; \bigl ( *\; \tilde V_{k} \bigr ) = (-1)^k\; \tilde V_{k} 
\end{array}\eqno(2.25)
$$
where $*$ operation corresponds to transformations in (2.10) and
$k = 0, 1, 2$ stands for the degree of the forms on the 2D compact manifold.
Another interesting point to be noted is the fact that {\it the requirement:
$I_{k} \rightarrow J_{k}$ under $*$ operation, singles out transformations
(2.10) from (2.10) and (2.11) which are symmetry transformations for the
 Lagrangian density (2.1)}. This assertion will play an important role in
the discussion of topological invariants and their 
transformations under $*$ operation for the case 
of non-Abelian gauge theory (see, e.g., Sec. 3 below).

To conclude this section, we note that 2D free $U(1)$ gauge theory is a 
prototype example of a field theoretical model for the Hodge theory. This
theory also turns out to be a new type of topological field theory. The form of
its Lagrangian density looks like Witten type of TFTs but its symmetries
are just like that of Schwarz type TFTs (as there is a conspicuous absence
of the topological shift symmetry in the theory).\\ 

\noindent
{\bf 3 Non-Abelian gauge theory}\\

\noindent
Let us start off with  the $(1 + 1)$-dimensional BRST invariant Lagrangian
density $({\cal L}_{B})$ for the self-interacting non-Abelian gauge theory 
in the Feynman gauge [26-28]
$$
\begin{array}{lcl}
{\cal L}_{B} &=& - \frac{1}{4}\; F^{\mu\nu a} F_{\mu\nu}^a 
- \frac{1}{2} (\partial \cdot A)^a (\partial \cdot A)^a  
- i \partial_{\mu} \bar C^a
D^{\mu} C^a \nonumber\\
 &=&  \frac{1}{2}\; E^a\; E^a 
- \frac{1}{2} (\partial \cdot A)^a (\partial \cdot A)^a  
- i \partial_{\mu} \bar C^a D^{\mu} C^a 
\end{array}\eqno (3.1)
$$
where $F_{01}^a = \partial_{0} A_{1}^a - \partial_{1} A_{0}^a
+ g f^{abc} A_{0}^b A_{1}^c = E^a $ is the ``coloured'' (group-valued) 
electric field derived from the ``coloured'' gauge connections 
$A_{0}^a$ and $A_{1}^a$,
$(\bar C^a) C^a$ are the anticommuting Faddeev-Popov (anti)ghost fields 
($ (\bar C^a)^2 = (C^a)^2 = 0$), the covariant derivative is:
$D_{\mu} C^a = \partial_{\mu} C^a + g \; f^{abc} A_{\mu}^b C^c$, spacetime
indices are: $\mu, \nu, \lambda.....= 0, 1$ and group indices
$ a, b, c....= 1, 2, 3...$ correspond to a compact Lie group, $g$ is the
coupling constant denoting the strength of the interaction amongst gauge
fields and structure constants $f^{abc}$ are chosen to be totally
antisymmetric for the above compact Lie algebra [33]. It has been demonstrated
[25] that the above Lagrangian density remains quasi-invariant
($\delta_{B} {\cal L}_{B} = - \eta \partial_{\mu} [(\partial \cdot A)^a
D^\mu c^a],\; \delta_{D} {\cal L}_{B} = \eta \partial_{\mu} [ E^a 
\partial^\mu \bar C^a]$) under the following on-shell ($\partial_{\mu}
D^\mu C^a = 0, \; D_{\mu} \partial^\mu \bar C^a = 0$) nilpotent
$(\delta_{B}^2 = \delta_{D}^2 = 0)$ BRST- and dual BRST transformations
$$
\begin{array}{lcl}
\delta_{B} A_{\mu}^a &=& \eta\; D_{\mu}\; C^a \;\;\;\;\;\qquad\;\;\;\;\;
\delta_{D} A_{\mu}^a = - \eta \varepsilon_{\mu\nu} \partial^{\nu} \bar C^a
\nonumber\\
\delta_{B} C^a &=& - \frac{\eta g}{2}
f^{abc} C^b C^c \;\qquad \;
\delta_{D} \bar C^a  = 0 \nonumber\\
\delta_{B} \bar C^a &=& - i \;\eta\; (\partial \cdot A)^a \;\; \qquad \;
\delta_{D} C^a = - i \;\eta \;E^a \nonumber\\
\delta_{B} E^a &=& \eta g f^{abc} E^b C^c\;\;\;\qquad \;\;
\delta_{D} (\partial \cdot A)^a  = 0 \nonumber\\
\delta_{B} (\partial \cdot A)^a &=& \eta \partial_{\mu} D^\mu C^a
\;\;\;\qquad \;\;\;\;\;
\delta_{D} E^a  = \eta D_{\mu} \partial^\mu \bar C^a 
\end{array}\eqno (3.2)
$$
where $\eta$ is an anticommuting ($\eta C^a = - C^a \eta, \; \eta \bar C^a
= - \bar C^a \eta)$ spacetime independent transformation parameter. It is
quite straightforward to check that the anticommutator of the above
transformations generate a bosonic ($\kappa = - i \eta \eta^\prime$)
symmetry transformation $(\delta_{W} = \{ \delta_{B}, \delta_{D} \})$
$$
\begin{array}{lcl}
\delta_{W} A_{\mu}^a &=& \kappa \;\bigl [\;D_{\mu} E^a
- \varepsilon_{\mu\nu} \partial^{\nu} (\partial \cdot A)^a
- i g f^{abc} \varepsilon_{\mu\nu} \partial^\nu \bar C^b C^c \;\bigr ]
\nonumber\\
\delta_{W} (\partial \cdot A)^a &=& \kappa\; \bigl [ \;\partial_{\mu}
D^\mu E^a + i g f^{abc} \varepsilon^{\mu\nu} \partial_{\mu} \bar C^b 
\partial_{\nu} C^c \;\bigr ] \nonumber\\
\delta_{W} E^a &=&  \kappa \;\bigl [ 
D_{\mu} \partial^\mu (\partial \cdot A)^a
- \varepsilon^{\mu\nu} D_{\mu} D_{\nu} E^a
+ i g f^{abc} D_{\mu} (\partial^\mu \bar C^b C^c)
\;\bigr ] \nonumber\\
\delta_{W} C^a &=& \;0 \;\;\;\qquad\;\;\;\; \delta_{W} \bar C^a = \;0
\end{array}\eqno(3.3)
$$
as the Lagrangian density ${\cal L}_{B}$ transforms to a total derivative [25]
$$
\begin{array}{lcl}
\delta_{W} {\cal L}_{B} &=& \kappa\; \partial_{\mu}\; \bigl [\; Z^\mu \;
\bigr ] \nonumber\\
Z^\mu &=& E^a \partial^\mu (\partial \cdot A)^a 
- (\partial \cdot A)^a D^\mu E^a + i g f^{abc}
( E^a \partial^{\mu} \bar C^b - \varepsilon^{\mu\nu} 
\partial_{\nu} \bar C^a (\partial \cdot A)^b ) C^c.
\end{array}\eqno (3.4)
$$
This bosonic symmetry transformation and the nilpotent symmetry 
transformations in (3.2) are generated by the conserved charges
$$
\begin{array}{lcl}
W^{(N)} &=& {\displaystyle \int}\; dx\;
\bigl [\;(\partial \cdot A)^a D_{0} E^a - E^a \partial_{0} (\partial 
\cdot A)^a - i g f^{abc} ( E^a \dot {\bar C^a} + \partial_{1} 
\bar C^a (\partial \cdot A)^b ) C^c \bigr ] \nonumber\\
Q_{B} &=& {\displaystyle \int}\; dx\;
\bigl [\;\partial_{0} (\partial \cdot A)^a C^a - (\partial \cdot A)^a
D_{0} C^a + \frac{ig}{2} f^{abc} \dot {\bar C^a} C^b C^c \;\bigr ] \nonumber\\
Q_{D} &=& {\displaystyle \int}\; dx\;
\bigl [\;E^a \dot {\bar C^a} - 
D_{0} E^a \bar C^a - ig f^{abc} \bar C^a 
\partial_{1} \bar C^b C^c \;\bigr ].
\end{array}\eqno (3.5)
$$
The continuous global scale invariance of the 
Lagrangian density (3.1) under transformations
$ C^a \rightarrow e^{-\Lambda} C^a,\; \bar C^a 
\rightarrow e^{\Lambda} \bar C^a,\;
A_{\mu}^a \rightarrow A_{\mu}^a$ (where $\Lambda$ is a global parameter)
leads to the derivation of a conserved ghost charge $(Q_{G})$
$$
\begin{array}{lcl}
Q_{G} = - i {\displaystyle \int}\; dx\;
\bigl [\;C^a \partial_{0}  \bar C^a +
\bar C^a D_{0} C^a \bigr ].
\end{array}\eqno (3.6)
$$
Together these generators obey the following extended BRST algebra
$$
\begin{array}{lcl}
&& Q_{B}^2 = \frac{1}{2} \{ Q_{B}, Q_{B} \} = 0
\qquad \;Q_{D}^2 = \frac{1}{2}\; \{ Q_{D}, Q_{D} \} = 0
\nonumber\\
&& \{ Q_{B}, Q_{D} \} =  W^{(N)} \qquad
 [ W^{(N)} , Q_{k} ] = 0 \;\;\;\qquad\;\;\; k = G, B, D \nonumber\\
&& i [ Q_{G}, Q_{B} ] = Q_{B} \qquad 
 i [ Q_{G}, Q_{D} ] = - Q_{D}.
\end{array}\eqno(3.7)
$$
This algebra is reminiscent of the algebra obeyed by the de Rham cohomology
operators $d, \delta , \Delta$ as given in (2.9a). As a consequence of this 
algebra, it is clear that, given a state $|\psi>_{n}$ with ghost number $n$
(i.e. $i Q_{G} |\psi>_{n} = n |\psi>_{n}$) in the quantum Hilbert space of
states, the following relations are correct
$$
\begin{array}{lcl}
i Q_{G} Q_{(B, D)} | \psi >_{n} = (n \pm 1) Q_{(B, D)} |\psi>_{n} 
\qquad  i Q_{G} W^{(N)}   | \psi >_{n} = n     W^{(N)}  |\psi>_{n}.
\end{array}\eqno(3.8)
$$   
This demonstrates that the ghost numbers for the states  $Q_{B} |\psi>_{n},
Q_{D} |\psi>_{n}$ and $W^{(N)} |\psi>_{n}$ are $(n+1), (n-1)$ and $n$
respectively. This is analogous to the change in the degree of the
form $f_{n}$, when operated upon by the cohomological operators
$d , \delta, \Delta$ defined on a compact manifold. It is now obvious that
the Hodge decomposition theorem (1.3) can be implemented in the quantum Hilbert
space of states $|\psi>_{n} = |\tilde \omega>_{n} + Q_{B} |\tilde \theta>_{n-1}
+ Q_{D} |\tilde \chi>_{n+1}$ (which is the analogue of (2.9b) for the
non-Abelian case).

It is well-known that, for the Witten type TFTs, the Lagrangian density
(or the action itself) is a BRST (anti)commutator (1.1).  With this as a 
backdrop, it can be noticed that, modulo some total derivatives, the 
Lagrangian density (3.1) can be recast as the sum of a BRST- and co-BRST 
anticommutator (or a BRST and co-BRST invariant parts)
$$
\begin{array}{lcl}
{\cal L}_{B} &=& \{ Q_{D}, S_{1}\} + \{ Q_{B}, S_{2}\} \nonumber\\
\eta {\cal L}_{B} &=& \delta_{D} \;(i S_{1}) + \delta_{B}\; (i S_{2}) \quad
S_{1} = \frac{1}{2} E^a C^a \quad S_{2} = - \frac{1}{2}
(\partial \cdot A)^a \bar C^a.
\end{array}\eqno (3.9)
$$
More precisely, the above expression can be seen to produce 
$$
\begin{array}{lcl}
{\cal L}_{B} &=& \frac{1}{2} E^a E^a - \frac{1}{2} (\partial \cdot A)^a
(\partial \cdot A)^a - i \partial_{\mu} \bar C^a D^\mu C^a
+ \partial_{\mu} \bigl [ X^\mu \bigr ] \nonumber\\
X^\mu &=& \frac{i}{2}  \bigl ( \bar C^a D^\mu C^a + \partial^\mu \bar C^a
C^a \bigr ). 
\end{array}\eqno(3.10)
$$
This shows that, with two nilpotent charges $Q_{B}$ and $Q_{D}$, the
Lagrangian density (3.1) resembles with that of the Witten type TFTs 
if we choose
the physical states as BRST- and co-BRST invariant (harmonic state) in the
Hodge decomposition theorem (i.e. $Q_{B}|phys> = 0, \; Q_{D} |phys> = 0$). It
will be noticed, however, that in our discussions of the symmetries for the
theory, we do not have any topological shift symmetry. Thus, from the symmetry
angle, 2D self-interacting non-Abelian gauge theory 
is like Schwarz type TFTs which possess only local gauge symmetries.

The expression for the symmetric energy-momentum tensor ($\tilde 
T_{\mu\nu}^{(s)}$) for the Lagrangian density (3.1) is
$$
\begin{array}{lcl}
\tilde T^{(s)}_{\mu\nu} &=&  
- \frac{1}{2}\; [\; \varepsilon_{\mu \rho} E^a + \eta_{\mu \rho}
(\partial \cdot A)^a\; ]\; \partial_{\nu} A^{\rho a}
- \frac{1}{2} \;[\; \varepsilon_{\nu \rho} E^a + \eta_{\nu \rho}
(\partial \cdot A)^a\; ]\; \partial_{\mu} A^{\rho a} \nonumber\\
&-& \frac {i}{2} (\partial_{\mu} \bar C^a) (\partial_{\nu} C^a 
+ D_{\nu} C^a) 
- \frac {i}{2} (\partial_{\nu} \bar C^a) (\partial_{\mu} C^a
+ D_{\mu} C^a) - \eta_{\mu \nu} \;{\cal L}_{B}.
\end{array}\eqno(3.11)
$$
Here ${\cal L}_{B}$ is the Lagrangian in (3.1) (or equivalently (3.9)). This
expression can be re-written, modulo some total derivatives, as the sum of
BRST- and co-BRST (anti)commutators (or, equivalently, as the BRST- and 
co-BRST invariant parts) 
$$
\begin{array}{lcl}
\tilde T^{(s)}_{\mu\nu} &=& \{ Q_{B}, L^{(1)}_{\mu\nu} \} 
+ \{ Q_{D}, L^{(2)}_{\mu\nu} \} \quad
\eta \tilde T^{(s)}_{\mu\nu} = \delta_{B} \bigl ( i L^{(1)}_{\mu\nu} \bigr )
+ \delta_{D} \bigl ( i L^{(2)}_{\mu\nu} \bigr )
\nonumber\\
L^{(1)}_{\mu\nu} &=& \frac{1}{2} \bigl [
(\partial_{\mu} \bar C^a) A_{\nu}^a +
(\partial_{\nu} \bar C^a) A_{\mu}^a + \eta_{\mu\nu} (\partial \cdot A)^a
\bar C^a \bigr ] \nonumber\\
L^{(2)}_{\mu\nu} &=& \frac{1}{2} \bigl [
(\partial_{\mu} C^a) \varepsilon_{\nu\rho} A^{\rho a} +
(\partial_{\nu} C^a) \varepsilon_{\nu\rho} A^{\rho a} 
- \eta_{\mu\nu} E^a C^a \bigr ].
\end{array}\eqno(3.12)
$$
For aesthetic reasons, we choose vacuum as well as physical states of the 
theory to be the harmonic states in the Hodge decomposition theorem because
they are BRST- and co-BRST invariant together (i.e. $ Q_{(B, D)} |vac> = 0,\;
Q_{(B, D)} |phys> = 0$). This will also ensure that there are no energy
excitations in the theory because the VEV (i.e. $< vac | \hat {\tilde T_{00}}
| vac > = 0$) as well as the excitations in the physical states (i.e.
$<phys | \hat {\tilde T_{00}} |phys^\prime> = 0$) turns out to be zero. 
This result reconfirms the topological nature of the theory under discussion.

Besides BRST- and co-BRST charges, there are anti- BRST and anti-dual BRST
charges in the theory which are also nilpotent of order two. For the 
non-Abelian gauge theories, the corresponding symmetries can be obtained only 
by introducing some auxiliary fields. The ensuing modified Lagrangian 
densities (which are equivalent extensions of (3.1)) are
$$
\begin{array}{lcl}
{\cal L_{B}} =  {\cal B}^a E^a - \frac{1}{2} \; {\cal B}^a {\cal B}^a
+ B^a (\partial \cdot A)^a + \frac{1}{2} (B^a B^a + \bar B^a \bar B^a) 
- i \partial_{\mu} \bar C^a D^{\mu} C^a
\end{array}\eqno (3.13a)
$$
$$
\begin{array}{lcl}
{\cal L}_{\bar B} =  {\cal B}^a E^a - \frac{1}{2} \; {\cal B}^a {\cal B}^a
- \bar B^a (\partial \cdot A)^a + \frac{1}{2} (B^a B^a + \bar B^a \bar B^a) 
- i D_{\mu} \bar C^a \partial^{\mu} C^a
\end{array}\eqno (3.13b)
$$
where ${\cal B}^a, B^a$ and $\bar B^a$ are the auxiliary fields. The latter 
two are restricted to satisfy the following relation [34]
$$
\begin{array}{lcl}
B^a + \bar B^a = i \;g \; f^{abc} \;C^b \;\bar C^c.
\end{array}\eqno (3.14)
$$
The following off-shell nilpotent ($ \delta_{AD}^2 = 0,\;
\delta_{AB}^2 = 0$) anti-BRST ($\delta_{AB}$)
and anti-dual BRST ($\delta_{AD}$) symmetry transformations
$$
\begin{array}{lcl}
\delta_{AB} A_{\mu}^a &=& \eta D_{\mu} \bar C^a \quad
\delta_{AB} \bar C^a = - \frac{\eta g}{2} f^{abc} \bar C^b \bar C^c \quad
\delta_{AB} C^a = i \eta \bar B^a \quad \delta_{AB} \bar B^a = 0 \nonumber\\
\delta_{AB} B^a &=& \eta g f^{abc} B^b \bar C^c \quad
\delta_{AB} {\cal B}^a = \eta g f^{abc} {\cal B}^b \bar C^c \quad
\delta_{AB} E^a = \eta g f^{abc} E^b \bar C^c
\end{array}\eqno (3.15a)
$$
$$
\begin{array}{lcl}
\delta_{AD} A_{\mu}^a &=& - \eta \varepsilon_{\mu\nu} \partial^\nu C^a \quad
\delta_{AD}  C^a = 0 \quad
\delta_{AD} \bar C^a = i \eta {\cal B}^a \quad 
\delta_{AD} \bar B^a = 0 \nonumber\\
\delta_{AD} B^a &=& 0 \qquad
\delta_{AD} {\cal B}^a = 0 \qquad
\delta_{AD} E^a = \eta D_{\mu} \partial^\mu C^a \qquad \delta_{AD}
(\partial \cdot A)^a = 0
\end{array}\eqno (3.15b)
$$
leave (3.13b) quasi invariant as:
$\delta_{AB} {\cal L}_{\bar B} = - \eta \partial_{\mu} [ \bar B^a D^\mu 
\bar C^a ], \; \delta_{AD} {\cal L}_{\bar B} 
=  \eta \partial_{\mu} [ {\cal B}^a \partial^\mu  C^a ]$. These symmetry 
transformations are generated by the following conserved charges
$$
\begin{array}{lcl}
Q_{AB} &=& {\displaystyle \int}\; dx\; \bigl [\;\dot {\bar B}^a 
\bar C^a - \bar B^a D_{0} \bar C^a - \frac{ig}{2} f^{abc} 
\dot C^a \bar C^b \bar C^c \;\bigr ] \nonumber\\
Q_{AD} &=& {\displaystyle \int}\; dx\;
\bigl [\;{\cal B}^a \dot  C^a - 
D_{0} {\cal B}^a  C^a - i g f^{abc}  C^a 
\partial_{1}  C^b \bar C^c \;\bigr ].
\end{array}\eqno (3.16)
$$
These charges further extend the BRST algebra (3.7) as given below
$$
\begin{array}{lcl}
&& Q_{AB}^2 = 0 \quad Q_{AD}^2 = 0 \quad [ W^{(N)}, Q_{AB} ] =
[ W^{(N)}, Q_{AD} ] = 0 
\quad \{ Q_{D}, Q_{AB} \} = 0 \nonumber\\
&& \{ Q_{AB}, Q_{AD} \} = W^{(N)} \quad \{Q_{B}, Q_{AB} \} = 0 \quad
\{Q_{D}, Q_{AD} \} = 0 
\nonumber\\
&& \{Q_{B}, Q_{AD} \} = 0 \quad i [ Q_{G}, Q_{AB} ] = - Q_{AB}
\quad i [ Q_{G}, Q_{AD} ] = Q_{AD}.
\end{array}\eqno (3.17)
$$
This algebra is also analogous to the algebra satisfied by the de Rham
cohomology operators (2.9a). Thus, we notice that the set
($Q_{AD}, Q_{AB}, W^{(N)}$) can also be identified with the set 
of cohomological operators ($d, \delta, \Delta$) defined on a compact manifold.

It is obvious that we have {\it four} conserved and nilpotent charges in
the theory. On a 2D compact manifold, topological invariants for the 
non-Abelian gauge theory can be defined analogous to the Abelian
gauge theory (cf.  (2.20)) by replacing $V_{k}$ and $W_{k}$ by $B_{k}, D_{k}$
and $A_{k}^{(b)}, A_{k}^{(d)}$. In fact, we can obtain a set of {\it three} 
topological invariants
($ B_{k}, k = 0, 1, 2$) w.r.t. the nilpotent and conserved BRST charge $Q_{B}$ 
for the Lagrangian density (3.13a) as 
\footnote{ Note that, in addition to the transformations in (3.2), there are
some off-shell nilpotent BRST transformations:
$ \delta_{B} \bar C^a = i \eta B^a\; \delta_{B} B^a = 0\;
\delta_{B} {\cal B}^a = \eta g f^{abc} {\cal B}^b C^c \;
\delta_{B} \bar B^a = \eta g f^{abc} \bar B^b C^c$ for (3.13a).}
$$
\begin{array}{lcl}
B_{0} &=& B^a C^a - \frac{ig}{2} f^{abc} \bar C^a C^b C^c \nonumber\\
B_{1} &=& \bigl [\; B^a A_{\mu}^a + i C^a D_{\mu} \bar C^a \;\bigr ]\; dx^\mu
\nonumber\\
B_{2} &=& i \; \bigl [\;A_\mu^a D_\nu \bar C^a - \bar C^a
D_{\mu} A_{\nu}^a \;\bigr ]\;
dx^\mu \wedge dx^\nu. 
\end{array}\eqno(3.18)
$$
Similarly, the set of three topological invariants 
($D_{k}, k = 0, 1, 2$) on the 2D compact manifold, defined
w.r.t. dual BRST charge $Q_{D}$, is 
$$
\begin{array}{lcl}
D_{0} &=& {\cal B}^a \bar C^a  \nonumber\\
D_{1} &=& \bigl [\; \bar C^a \varepsilon_{\mu\rho} \partial^\rho C^a 
- i {\cal B}^a A_{\mu}^a \; \bigr ]\; dx^\mu \nonumber\\
D_{2} &=& i\; \bigl [\;\varepsilon_{\mu\rho} \partial^\rho C^a A_\nu^a
+ \frac{1}{2} C^a \varepsilon_{\mu\nu} (\partial \cdot A)^a \;\bigr ]\;
dx^\mu \wedge dx^\nu. 
\end{array}\eqno(3.19)
$$
The definition of the anti- BRST charge ($Q_{AB}$) in non-Abelian gauge theory
is more involved as we introduce some new auxiliary fields. In terms of these
auxiliary fields and the basic fields, the three invariants ($A^{(b)}_{k},
k = 0, 1, 2$) corresponding to zero-, one- and two-forms are
$$
\begin{array}{lcl}
A^{(b)}_{0} &=& \bar B^a \bar C^a 
- \frac{ig}{2} f^{abc} C^a \bar C^b \bar C^c \nonumber\\
A^{(b)}_{1} &=& \bigl [\; \bar B^a A_{\mu}^a + i \bar C^a D_{\mu}  C^a 
\;\bigr ]\; dx^\mu \nonumber\\
A^{(b)}_{2} &=& i \;\bigl [ \; A_\mu^a D_\nu C^a -  C^a
D_{\mu} A_{\nu}^a \; \bigr ] \;
dx^\mu \wedge dx^\nu. 
\end{array}\eqno(3.20)
$$
Finally, three invariants ($A^{(d)}_{k}, k = 0, 1, 2$), corresponding to 
the anti-dual BRST charge ($Q_{AD}$), are
$$
\begin{array}{lcl}
A^{(d)}_{0} &=& {\cal B}^a  C^a  \nonumber\\
A^{(d)}_{1} &=& \bigl [\; C^a (\varepsilon_{\mu\rho} \partial^\rho \bar C^a ) 
- i {\cal B}^a A_{\mu}^a \; \bigr ] \;dx^\mu \nonumber\\
A^{(d)}_{2} &=&  i \;\bigl [ \;\varepsilon_{\mu\rho} \partial^\rho \bar C^a 
A_\nu^a + \frac{1}{2} \bar C^a \varepsilon_{\mu\nu} 
(\partial \cdot A)^a \;\bigr ] \; dx^\mu \wedge dx^\nu. 
\end{array}\eqno(3.21)
$$
These topological invariants obey the same type of recursion relations as
are expected of the topological invariants of a well-defined TFT. These 
relations are [2,4,35]
$$
\begin{array}{lcl}
\delta_{B} \;B_{k} &=& \eta \;d \; B_{k -1} \qquad
\delta_{AB} \;A^{(b)}_{k} = \eta\; d \; A^{(b)}_{k -1} \qquad
d = dx^\mu \; \partial_{\mu} \nonumber\\
\delta_{D} \;D_{k} &=& \eta \;\delta\; D_{k-1} 
\qquad \delta_{AD}\; A^{(d)}_{k-1} = \eta \;\delta \;A^{(d)}_{k}
\qquad \;\delta = i \;dx^\mu \; \varepsilon_{\mu\nu} \partial^{\nu}. 
\end{array}\eqno(3.22)
$$
These properties establish the topological nature of 2D self-interacting 
non-Abelian gauge
theory. This theory belongs to a new class of TFT as is evident from its
differences with Witten- and Schwarz type of TFTs.

As was emphasized at the end of equation (2.25), one can obtain an analogue
of the $(*)$ operation from the requirement that the topological invariants
of the self-interacting non-Abelian gauge theory should obey the same kind
of relations as (2.24) and (2.25) for the Abelian gauge theory. It can be
checked that under the following transformations
$$
\begin{array}{lcl}
&&C^a \rightarrow i \bar C^a \quad \bar C^a \rightarrow i C^a
\quad A_{\mu}^a \rightarrow A_{\mu}^a \nonumber\\
&& B^a \rightarrow - i {\cal B}^a - \frac {ig}{2} f^{abc} C^b \bar C^c
\quad {\cal B}^a \rightarrow - i  B^a - \frac {g}{2} f^{abc} C^b \bar C^c
\nonumber\\
&& \partial_{\mu} \delta^{ab} \rightarrow
i \varepsilon_{\mu\rho} \partial^\rho \delta^{ab} +
\frac {g}{2} f^{abc} A_{\mu}^c \quad
 \varepsilon_{\mu\rho} \partial^\rho \delta^{ab} \rightarrow i \partial_\mu
\delta^{ab} - \frac {ig}{2} f^{abc} A_{\mu}^c \nonumber\\
\end{array}\eqno(3.23)
$$
the topological invariants, for $ k = 0, 1, 2$, transform as
$$
\begin{array}{lcl}
 B_{k} \rightarrow D_{k}\; \quad
A_{k}^{(b)} \rightarrow A_{k}^{(d)} \quad  
D_{k} \rightarrow (-1)^k \; B_{k} 
\quad  A_{k}^{(d)} \rightarrow (-1)^k\; A_{k}^{(b)}.
\end{array}\eqno(3.24)
$$
In the proof of the above relations, we should include in (3.23)
$$
\begin{array}{lcl}
{\cal B}^a &\rightarrow& - i \bar B^a - \frac{g}{2}\;f^{abc} \bar C^b C^c
\nonumber\\
\bar B^a &\rightarrow& - i {\cal B}^a - \frac{ig}{2}\;f^{abc} \bar C^b C^c
\end{array}\eqno(3.25)
$$
for the checking of the transformation 
properties of topological invariants w.r.t. $Q_{AB}$ and
$Q_{AD}$. In the language of the above $(*)$ operation, the relations 
(3.24) can be mathematically expressed as
$$
\begin{array}{lcl}
*\; B_{k} = D_{k} \quad * \;A_{k}^{(b)} = A_{k}^{(d)} \quad
*\; ( *\; B_{k}\;) = (-1)^k\; B_{k} \quad
*\; ( *\; A_{k}^{(b)}\;) = (-1)^k\; A_{k}^{(b)}
\end{array}\eqno(3.26)
$$
It will be noticed that (3.24, 3.26)) are exactly like (2.24, 2.25). 
However, the
transformations (3.23, 3.25) are still not the analogue of the exact Hodge dual
$(*)$ operation of differential geometry. Unlike the case of 2D Abelian gauge
theory where this analogy was perfect, we see that, for the non-Abelian case,
the above transformations do not keep the Lagrangian density (3.1)
or (3.13a,b) invariant. 
Furthermore, the dual BRST transformation $\delta_{D}$ can not be obtained
from BRST transformations $\delta_{B}$ {\it by exploiting (3.23)}. In addition,
the analogues of (2.14) and (2.16) do not exist.

We conclude this section with the remark that self-interacting 2D non-Abelian
gauge theory is a topological field theory which bears the appearance of a 
Witten type TFT but possesses symmetries that are reminiscent of a 
Schwarz type of TFT. \\

\noindent
{\bf 4 Summary and discussion}\\

\noindent
It has been shown that free Abelian- and self-interacting 
non-Abelian gauge theories in 2D belong to a new class of topological field
theories. The ideas of BRST cohomology and Hodge decomposition theorem play
a pivotal role in the proof of topological nature of these theories. The
{\it local} symmetries of these theories define the de Rham cohomology 
operators and a discrete symmetry transformation 
corresponds to the Hodge $(*)$ operation of differential geometry. 
As far as these symmetries are concerned, there 
are some  specific differences as well as similarities between 2D free
Abelian- and self-interacting
\footnote { Besides interaction among themselves, the non-Abelian gauge fields
also interact with ghost fields. The latter fields are, however, not the
physical matter fields.} non-Abelian gauge theories. For
instance, the curvature tensor for the Abelian gauge theory is derived
from $F^{(A)} = d A$ when {\it $d$ directly operates on the one-form $A$}.
This is not the case with the non-Abelian gauge theory where $F^{(N)}
= (d + A)\wedge A$. Under the BRST transformations, however, the kinetic
energy terms for Abelian- and non-Abelian gauge theories do remain invariant
{\it even though $\delta_{b} F^{(A)}_{\mu\nu} = 0$ 
but $\delta_{B} F^{a (N)}_{\mu\nu} = \eta g f^{abc} F_{\mu\nu}^{b (N)} C^c$}. 
Under the dual BRST symmetries, the
gauge-fixing terms of both the theories remain invariant as they are obtained
by {\it the application of $\delta = \pm * d *$ operator on the one-form $A$}. 
The Casimir operator generates a symmetry transformation in which ghost fields
do not transform and $A_{\mu}$ gauge field transforms to its own 
equation of motion in both the cases (cf. (2.4), (3.3)).

It is interesting to note that, in the Abelian gauge theory, the requirement
that the topological invariants should be related with each-other
($ I_{k} \rightarrow J_{k}$) by a Hodge $(*)$ operation, singles out (2.10)
from the set of transformations
(2.10) and (2.11). On the other hand, it can be seen that for
the {\it interacting} 2D Abelian gauge theory where Abelian gauge field
$A_{\mu}$ couples with the Dirac fields, it is (2.11) that is singled
out for the generalization to include matter (Dirac) fields [36] and (2.10)
is ruled out for such an important extension. 
Furthermore, it is gratifying to note that certain specific
transformation properties of the topological invariants (cf. (2.24), (2.25)
(3.24), (3.26)) lead to the derivation of transformations (3.23) for the
non-Abelian gauge theory which reduce to the Abelian case (2.10) under
the limit $ g \rightarrow 0$. In fact, it seems to us, the root cause
of the lack of a perfect definition of Hodge $(*)$ operation for the
non-Abelian gauge theory, in the language of symmetry property, is the
difference in the definition of a non-Abelian curvature tensor $F^{(N)} 
= (d + A) \wedge A$ from that of an Abelian gauge theory where $F^{(A)} 
= d A$. The latter is a perfect Hodge- as well as topological field theory. 
In fact, for the non-Abelian gauge theories, it can be seen that the set 
($ d + A, \delta, \Delta$) does not define the perfect de Rham 
cohomological operators whereas for the Abelian gauge theory the set 
($d, \delta, \Delta$) does define the same.

In some of our works [23-25,32] (including the present one), we have 
carried out investigations in the context of BRST cohomology and 
HDT in 2D and 4D where spacetime is endowed with a flat Minkowski
metric. The topological nature emerges due to the fact that there are no
propagating degrees of freedom associated with the gauge bosons. It would
be an interesting endeavour to consider the interacting gauge theories where
matter fields are also present. Some steps in this direction [36,37] 
have already
been taken. It appears to us, at the moment, that the BRST cohomology and
HDT would shed some light on the Adler-Bardeen-Jackiw (ABJ) anomaly in
2D where a $U(1)$ gauge field is coupled to the conserved current of
the Dirac fields. For such an interacting theory, it has been shown that
the dual BRST symmetry is connected with the chiral ($\gamma_{5}$) 
transformation on the matter (Dirac) fields [36,37]. It would be nice to
generalize this assertion to the non-Abelian gauge theory in 2D where there
is an interaction with matter (Dirac) fields.

The central outcome of of our studies of the BRST cohomology and HDT should 
be taken as the proof of the existence of a new class of topological field
theories which
are nothing but the free 2D Abelian- and self-interacting non-Abelian 
gauge theories. Furthermore, we strongly feel that our studies would
shed some light on the consistency and unitarity of the anomalous
gauge theories in 2D (see, e.g., Refs. [38,39] and references therein).
These studies might provide an insight to study thoroughly 
$(3 +1)$-dimensional TFTs in the framework of BRST cohomology and HDT which
will have something to do with the real spacetime manifolds.

The key results in the study of TFTs with a nontrivial spacetime metric
has been the classification of 3D and 4D manifolds which have been of
importance in the context of string theories [2,4,6,7]. It would be a nice
future direction to study 2D and 4D theories [23-25,32] with a nontrivial 
spacetime metric in the framework of BRST cohomology and HDT and explore
the outcome of such studies. In fact, we guess that the results of earlier
works [23-25,32] can be generalized to the case of nontrivial metrics. For
this to be true, one has to show the metric independence of the path integral 
measure. In Ref. [35], it has been established that
the existence of a BRST type fermionic-bosonic symmetry is good enough
to prove the metric independence of the measure. It is obvious that,
we have such kind of BRST- and co-BRST symmetries in our theories. 
Thus, heuristically, it appears to us that the measure will be
independent of the choice of the metric for these theories as well. After this,
it will be straightforward to show that the partition functions as well as
the expectation values of the BRST-, co-BRST- and topological invariants 
are metric independent. To demonstrate this, one has to require that:
$Q_{B} |phys> = 0, Q_{D} |phys> = 0$ (see, e.g., Ref. [12]). This 
requirement, very clearly, entails on the physical
states of the theory to be the harmonic state of the HDT.  These are some of 
the issues which are under investigation and our 
results will be reported elsewhere [40].

\end{document}